\documentclass[aps,preprint,prd,showpacs,nofootinbib]{revtex4}

\usepackage{latexsym}
\usepackage{amsmath}
\usepackage{graphicx}
\usepackage{subfigure}
\usepackage{dcolumn}
\usepackage{bm}
\usepackage{amssymb}
\usepackage{color}
\usepackage{float}
\usepackage[colorlinks,linkcolor=magenta,anchorcolor=cyan,citecolor=blue]{hyperref}

\hypersetup{colorlinks=true,
    breaklinks=true,
    pdfstartview=Fit,
    linkcolor=blue,
    citecolor=blue,
    urlcolor=blue}

\def\lf{\left}
\def\rt{\right}

\def\nn{\nonumber}
\def\be{\begin{equation}}
    \def\ee{\end{equation}}
\def\ba{\begin{eqnarray}}
    \def\ea{\end{eqnarray}}

\begin{document}

\title{Implication of island for inflation and primordial perturbations}

\author{Yun-Song Piao$^{1,2,3,4} $ \footnote{\href{yspiao@ucas.ac.cn}{yspiao@ucas.ac.cn}}}

    \affiliation{$^1$ School of Fundamental Physics and Mathematical
        Sciences, Hangzhou Institute for Advanced Study, UCAS, Hangzhou
        310024, China}

    \affiliation{$^2$ School of Physics Sciences, University of
        Chinese Academy of Sciences, Beijing 100049, China}

    \affiliation{$^3$ International Center for Theoretical Physics
        Asia-Pacific, Beijing/Hangzhou, China}

    \affiliation{$^4$ Institute of Theoretical Physics, Chinese
        Academy of Sciences, P.O. Box 2735, Beijing 100190, China}

    \begin{abstract}

It is usually thought that the efolds number of inflation must be
bounded by its de Sitter entropy, otherwise we will have an
information paradox. However, in light of the island rule for
computing the entanglement entropy, we show that such a bound
might be nonexistent, while the information flux of primordial
perturbation modes the observer after inflation is able to detect
follows a Page curve. In corresponding eternally inflating
spacetime, it seems that our slow-roll inflation patch is
accompanied with a neighbourly collapsed patch (eventually
developing into a black hole) so that its Hawking radiation might
be just our primordial perturbations. Accordingly, the
perturbation spectrum we observed will present a ``Page-like"
suppression at large scale.

    \end{abstract}

    \maketitle
    \tableofcontents

\section{Introduction}

The success of inflation
\cite{Guth:1980zm,Linde:1981mu,Albrecht:1982wi,Starobinsky:1980te,Linde:1983gd}
suggests the existence of a de Sitter (dS) phase in the very early
stage of our observable Universe. Generally, the inflation is
required to last $N_{efolds}\gtrsim 60$, where \be N_{efolds}=\ln
{k_e\over k}=\int Hdt\ee is the efolds number for primordial
perturbation before the inflation ended, $k$ is the comoving
wavenumber of perturbation mode and $k_e=a_eH$.

During inflation, the perturbation modes exit the horizon, and
become the primordial perturbations likely accessible to an
asymptotic observer inside the flat Minkowski-like patch after
inflation.
It is conjectured that such an observer will be able to access a
large number of modes and associate these perturbation modes to an
entropy: \be S_{mat}\sim \ln{k_e\over k}\simeq
N_{efolds}.\label{kek}\ee


Though dS space is infinite, it has a finite entropy
$S_{dS_4}\simeq {1/H^2}$ so that the post-inflation observer is
never able to observe independent perturbation modes more than
$e^{S_{dS_4}}$, i.e.$S_{mat}\lesssim S_{dS_4}$. As a result, it
has been argued in Ref.\cite{Arkani-Hamed:2007ryv} that else at
the semiclassical level the efolds number of inflation is bounded
by $S_{dS_4}$, \be N_{efolds}\lesssim S_{dS_4}, \label{dSbound}\ee
or we will have a black hole-like information paradox
\cite{Hawking:1976ra}, i.e. the information flux of perturbation
modes follows the Hawking curve, not the Page curve
\cite{Page:1993wv}.

It is well-known that recently a breakthrough
\cite{Penington:2019npb,Almheiri:2019psf,Almheiri:2019hni}
on the black hole information paradox has been made. It is thought
that at later stage of black hole, an ``island" (covering most of
black hole interior) will appear, and the Hawking radiation near
null infinity must be entangled with the state of such an island
so that the Page curve reflecting the information conservation can
be recovered.





Thus it is worth exploring whether such an ``island" could bring
us a different insight into the entropy bound for inflation
\cite{Almheiri:2019yqk}
(see e.g. recent \cite{Teresi:2021qff,Seo:2022ezk})
\footnote{Recently, the applications of island rule have been
studied intensively, e.g.
\cite{Almheiri:2019psy,Gautason:2020tmk,Alishahiha:2020qza,Li:2020ceg,Dong:2020uxp,Ling:2020laa,Matsuo:2020ypv,Wang:2021woy,Wang:2021mqq,Chu:2021gdb,Yu:2021cgi,Azarnia:2021uch,He:2021mst,Yu:2021rfg,Gan:2022jay,Miao:2022mdx,Choudhury:2020hil,Choudhury:2022mch},
which might also have interesting implications for inflation.},
and primordial perturbations and whether the information flux of
perturbation modes in our observable Universe actually follows a
Page curve or not, what it hints ?



In section-II, with a Jackiw-Teitelboim-like (JT) inflation, where
the dS patch near ${\cal I}^+$ is jointed to a Minkowski patch,
following Refs.\cite{Teresi:2021qff,Seo:2022ezk}, we briefly
review the evolution of primordial-perturbation-like CFT modes.

In section-III.A, in light of the island rule for computing the
entanglement entropy, we show that
the bound (\ref{dSbound}) might be nonexistent,
while the information flux of primordial perturbation modes
accessible to the asymptotic observer inside the Minkowski patch
follows a Page curve.

In section-III.B and C, we discuss its implication for inflation
and primordial perturbations. In corresponding eternally inflating
multiverse, it seems that our patch must be entangled with a
neighbourly collapsed patch (eventually developing into a black
hole) so that its Hawking radiation might be just the primordial
perturbations in our observable Universe. Accordingly, the
spectrum of primordial perturbations observed will be modified,
which might present a ``Page-like" suppression at
$N_{efolds}\gtrsim N_{Pagefolds}$ scale.



.


\section{A JT-like model of inflation }


\subsection{Primordial ``perturbations" and entropy accessible to an
asymptotic observer}

\begin{figure}[t]
\begin{center}
\includegraphics[width=14cm]{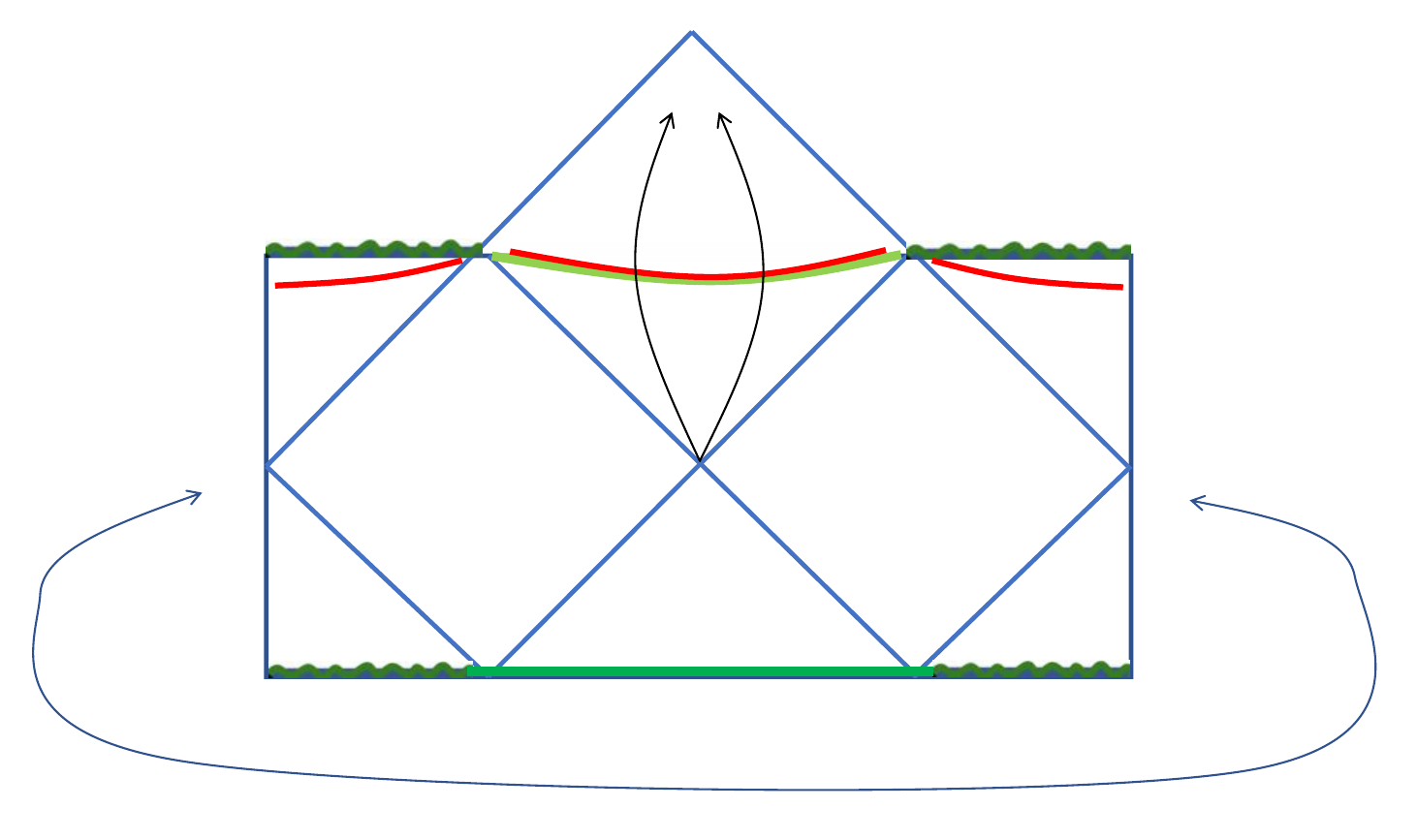}
\caption{A JT-like model of inflation. The dS patch is jointed to
a Minkowski hat at the ``reheating" surface (green curve), and $R$
(red curve) near which is entangled with the island $I$ (red
curve) inside the black hole. The black lines represent the
evolution of perturbation modes. } \label{Fig.1}
\end{center}
\end{figure}

In a JT-like model of inflation
\cite{Chen:2020tes,Hartman:2020khs}, see also
Refs.\cite{Teitelboim:1983ux,Jackiw:1984je,Maldacena:2019cbz,Cotler:2019nbi},
the dS patch near ${\cal I}^+$ is straightly jointed to a
Minkowski patch,
described as, \ba I &\sim & S_0\int R\sqrt{-g}d^2 x \nonumber\\
&&+\int \left(\phi R - 2V_{eff}\right)\sqrt{-g}d^2 x + I^{CFT}
\,\label{S}\ea where the boundary parts have been left out, and
$I^{CFT}$ is CFT.


In detail, such a JT inflation is explained as follows, see
Fig.\ref{Fig.1}. In global coordinates, for $V_{eff}=H^2\phi$, we
have \be ds^2 = {1\over H^2\cos^2 \sigma} \left(-d\sigma^2 +
d\theta^2\right),\label{SchdS}\ee and $\phi = \phi_r
({\cos\theta\over \cos\sigma})$ with $\phi_r>0$.
Here, the expanding dS patch responsible for inflation is
accompanied with the collapsed patches on its left and right side,
respectively, see Appendix A.
However, the big-bang-like evolution must start after a period of
inflation, which is implemented by straightly jointing the dS
patch to a flat Minkowski patch ($V_{eff}=1$) at $\phi\gg 1$, see
such a Minkowski hat in Fig.\ref{Fig.1}.

In the hyperbolic coordinates, for the expanding dS patch, we have
\be ds^2= \frac{1}{\sinh^2 (T/H^{-1})}\lf(-dT^2+dX^2\rt),
\quad \phi = -\phi_r \coth (T/H^{-1}),\label{dsTX}\ee where $T<0$.
Inside the collapsed patch, the metric is actually also
(\ref{dsTX}), but the dilaton $\phi$ has the opposite sign, see
also
Refs.\cite{Aguilar-Gutierrez:2021bns,Levine:2022wos,Balasubramanian:2020xqf,Yadav:2022jib,Baek:2022ozg}.

The observer (as a \textsf{cosmic census taker}
\cite{Susskind:2007pv,Sekino:2009kv}) inside the Minkowski hat
will ``see" a region $R$ ($-X_R\leqslant X\leqslant X_R$) at the
jointing surface. Thus such an observer will be able to access to
the primordial-perturbation-like CFT modes stretched by inflation.
Accordingly, the entropy associated with such perturbation modes
might be nothing but the entanglement entropy of CFT.

In 2D CFT the entanglement entropy of an interval with length $l$
is \be S=\frac {c} {6} \log
\left({l^2}/{\epsilon_{uv}^2}\right),\ee with $\epsilon_{uv}$ as
the cutoff scale. Thus the entanglement entropy at $R$ is (see
Appendix A) \be S_{mat}(R)= \frac {c} {3} \log \frac{2 \sinh
(X_R/H^{-1})}{H\epsilon_{uv}\sinh(-T_R/H^{-1})}, \label{SmatR}\ee
where the central charge of CFT must satisfy $1\ll c\ll \phi_r$ so
that the backreaction is negligible. This is the entanglement
entropy for the Hartle-Hawking (Bunch-Davies) state on dS$_2$.
(\ref{SmatR}) suggests that the entropy of perturbation modes an
observer inside the Minkowski hat could detect is $S_{mat} \sim
X_R$.


The perturbation mode (with the wavelength $X_R$) that the
observer inside the Minkowski hat see can be back to earlier
$|T|=X_R$. Thus \be S_{mat}(R) \approx \frac {c} {3} \log \frac{2
\sinh (-T/H^{-1})}{H\epsilon_{uv}\sinh(-T_R/H^{-1})}\sim \frac {c}
{3}\log{a_R\over a}. \label{SRefolds}\ee Here, \be
N_{efolds}=\ln{k_R\over k}=\ln{a_R\over a}.\ee
Thus we have \be S_{mat}(R)\thickapprox \lf(\frac {c}
{3}\rt)N_{efolds}.\ee

\section{Implication of island for inflation and primordial perturbations}

\subsection{de Sitter entropy bound ?}

\begin{figure}[t]
\begin{center}
\includegraphics[width=12cm]{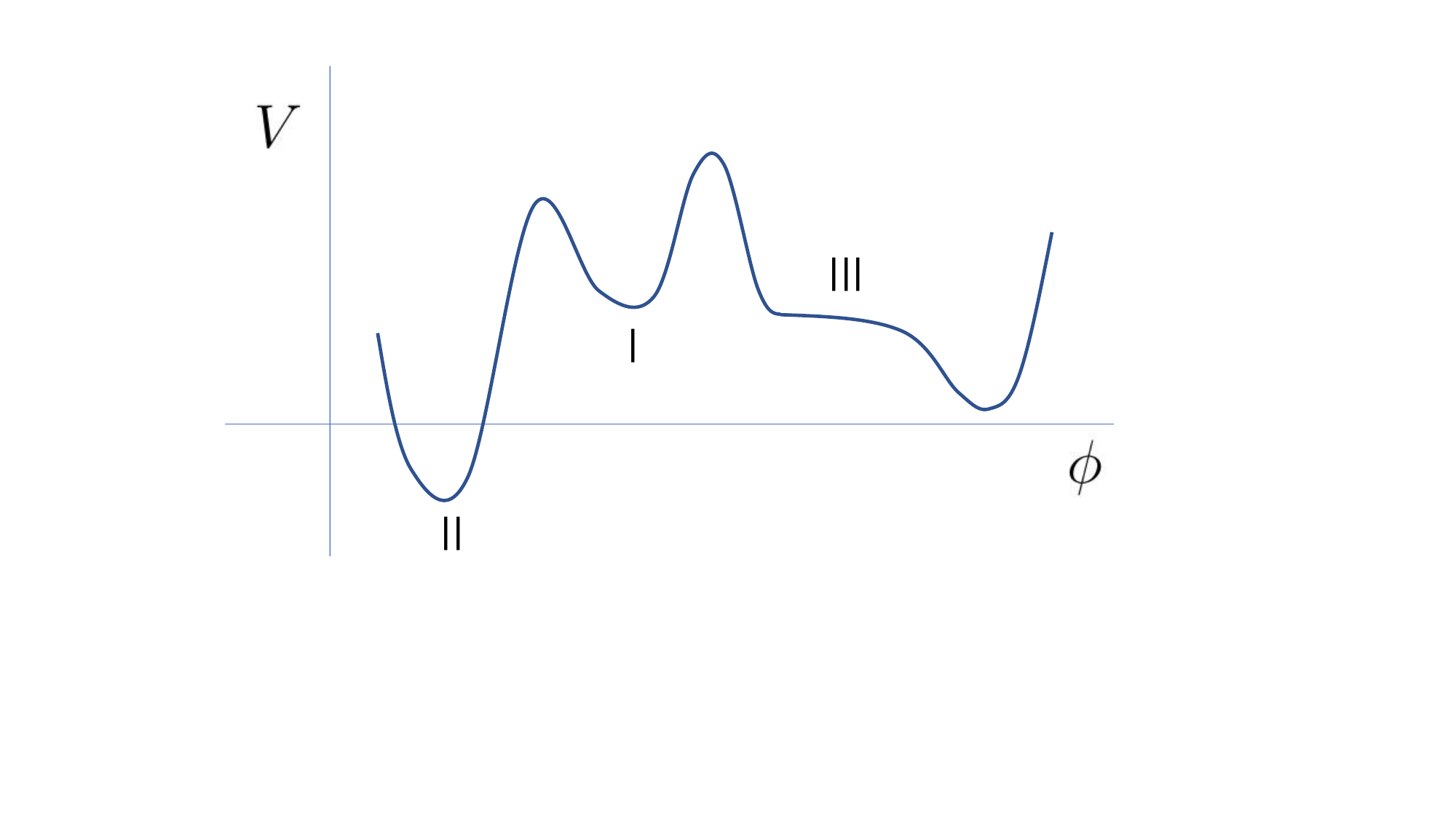}
\caption{The sketch of a simplified landscape. Initially, the
``universe" is at I, after a period some local regions will be
populated by the bubbles with different dS and AdS vacua. The
patch with III went through a period of slow-roll inflation and
will eventually evolve into our observable Universe, while the
patch with II will collapse into a black hole. } \label{Fig.3}
\end{center}
\end{figure}

\begin{figure}[t]
\begin{center}
\includegraphics[width=14cm]{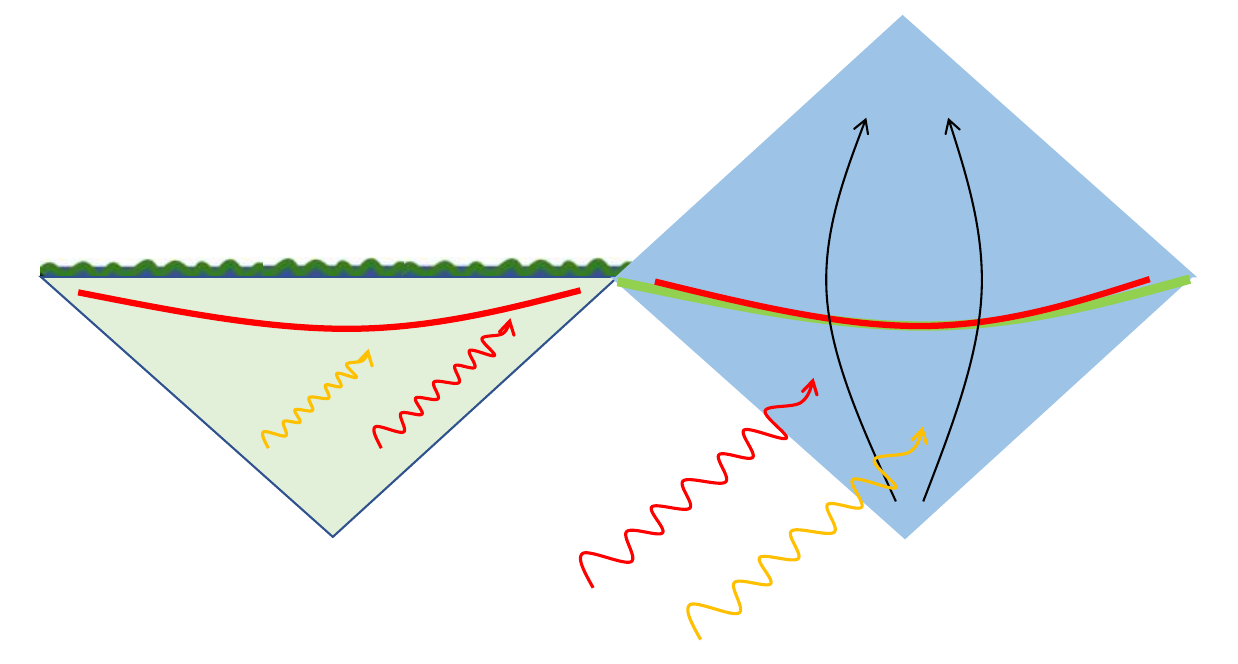}
\caption{ A local Fig.\ref{Fig.1}-like (JT-inflation-like) region
in an inflating
multiverse. 
The wavy lines are a pair of entangled partner, one falls into the
collapsing patch, and the other escapes into our slow-roll
inflating patch and will ``develop" into our primordial
perturbation. } \label{Fig.4}
\end{center}
\end{figure}

In a well-motivated landscape \cite{Bousso:2000xa,Kachru:2003aw},
both dS and anti-de Sitter (AdS) vacua might coexist. In the 4D
eternally inflating multiverse
\cite{Guth:2000ka,Vilenkin:2006xv,Linde:2014nna,Linde:2015edk},
different patches will be populated by different vacua. The dS or
AdS bubbles after nucleating will expand with light velocity, and
the coordinate radius of bubble will be rapidly asymptotic to the
horizon of parent dS region \cite{Guth:1982pn}, \be r=\int_{\ll
1/H}^{t\gg 1/H}{dt\over e^{Ht}}\simeq {1/H}.\ee The bubbles with a
period of slow-roll inflation will evolve into our observable
Universe, see Fig.\ref{Fig.3}), while the universe inside the AdS
bubbles will inevitably collapse \footnote{Though the bubble wall
of AdS bubble is expanding, the spacetime inside it is collapsing,
see Refs.\cite{Lin:2021ubu,Lin:2022ygd} for relativistic
simulation.}, e.g.\cite{Felder:2002jk}, so that a AdS bubble will
eventually develop into a black hole \footnote{However, an AdS
phase near the recombination might helps resolve the recent Hubble
tension, and so AdS vacua can have potential observable imprints
in CMB \cite{Ye:2020btb,Ye:2020oix,Jiang:2021bab,Ye:2021iwa}.}.
The mass of such a black hole is \be M_{BH}\simeq {4\pi\over 3
H^3} M_p^2 H^2= {M_p^2\over H}, \label{MBH}\ee so we approximately
have $R_{BH}\sim {1/H}$. Therefore, in such an eternally inflating
landscape, we not only have local slow-roll inflation patches but
also might also have some black holes, which is the scenario we
will consider.

Inside the patch (or bubble) with a period of slow-roll inflation,
by observing the density perturbations in the CMB, an observer
inside the Minkowski hat will be able to assign the state of such
perturbation modes to early dS phase, see Fig.\ref{Fig.4}. Thus
the maximal number of independent modes he detected should be
bounded by the dimensionality of the dS Hilbert space $\sim
e^{S_{dS_4}}$ \cite{Arkani-Hamed:2007ryv}. This suggests
$N_{efolds}\lesssim S_{dS_4}$, i.e.(\ref{dSbound}).

However, ``locally" \footnote{Here, we only focus on a local
region, i.e. a slow-roll inflation patch is accompanied with a
black hole, of eternal inflating spacetime. } such a scenario in
Fig.\ref{Fig.4} is similar to the 4D Schwarzschild-dS spacetime
but with $R_{BH}\simeq {1/H}$, so a $dS_2\times S^2$ region.
Thus it might be expected that the 2D CFT result for the entropy
of perturbation modes is also applicable for such a local
JT-inflation-like part in the inflating multiverse in
Fig.\ref{Fig.4}.


Here, 
regarding the primordial perturbation modes as 2D CFT-like (the
$S^2$ coordinates are neglected \footnote{see
e.g.Ref.\cite{Hashimoto:2020cas} for 4D Schwarzschild black
holes.}), the observer inside the Minkowski hat after slow-roll
inflation will be able to observe the perturbation modes at $R$
with \be S_{\textsf{ no-island}}(R)=S_{mat}(R)\simeq \lf({c\over
3}\rt)N_{efolds},\label{Sno}\ee which corresponds to (\ref{kek}).



However, it is possible that $R$ is entangled with an ``island"
(set at $T=T_I$, $-X_I \leqslant X\leqslant X_I$) inside the
neighbourly collapsing patch \footnote{see also
Refs.\cite{Aalsma:2021bit,Kames-King:2021etp} for recent
approaches without the collapsing patch.}. The entanglement
entropy with one endpoint at $I$ and other at $R$ is \be S_{ent}=
\frac c 6 \log \frac{2\cosh(\sum_{p=R,I}X_p/H^{-1}) + 2\cosh
(\sum_{p=R,I}T_p/H^{-1})}{\epsilon_{uv}^2 \prod_{p=R,I}sinh
(T_p/H^{-1})}. \label{Sent}\ee Thus we have  \ba
S_{gen}(R \cup I)&\sim & 2S_{dS_4} +2\phi_r coth (T_I/H^{-1})\nonumber\\
&+& \frac c 3 \log \left(\frac{2\cosh(\sum_{p=R,I}X_p/H^{-1}) +
2\cosh (\sum_{p=R,I}T_p/H^{-1})}{\epsilon_{uv} \epsilon_{rg}
\prod_{p=R,I}sinh (T_p/H^{-1})}\right), \label{Sgen}\ea where
$S_0+\phi_r coth (T_I/H^{-1})$ is that at the endpoint of island,
and $S_0\sim S_{dS_4}$ (see \cite{Maldacena:2019cbz} for the 4D
Schwarzschild-dS spacetime with $R_{BH}\simeq {1/H}$ to $dS_2$)

\begin{figure}[t]
\begin{center}
\includegraphics[width=15cm]{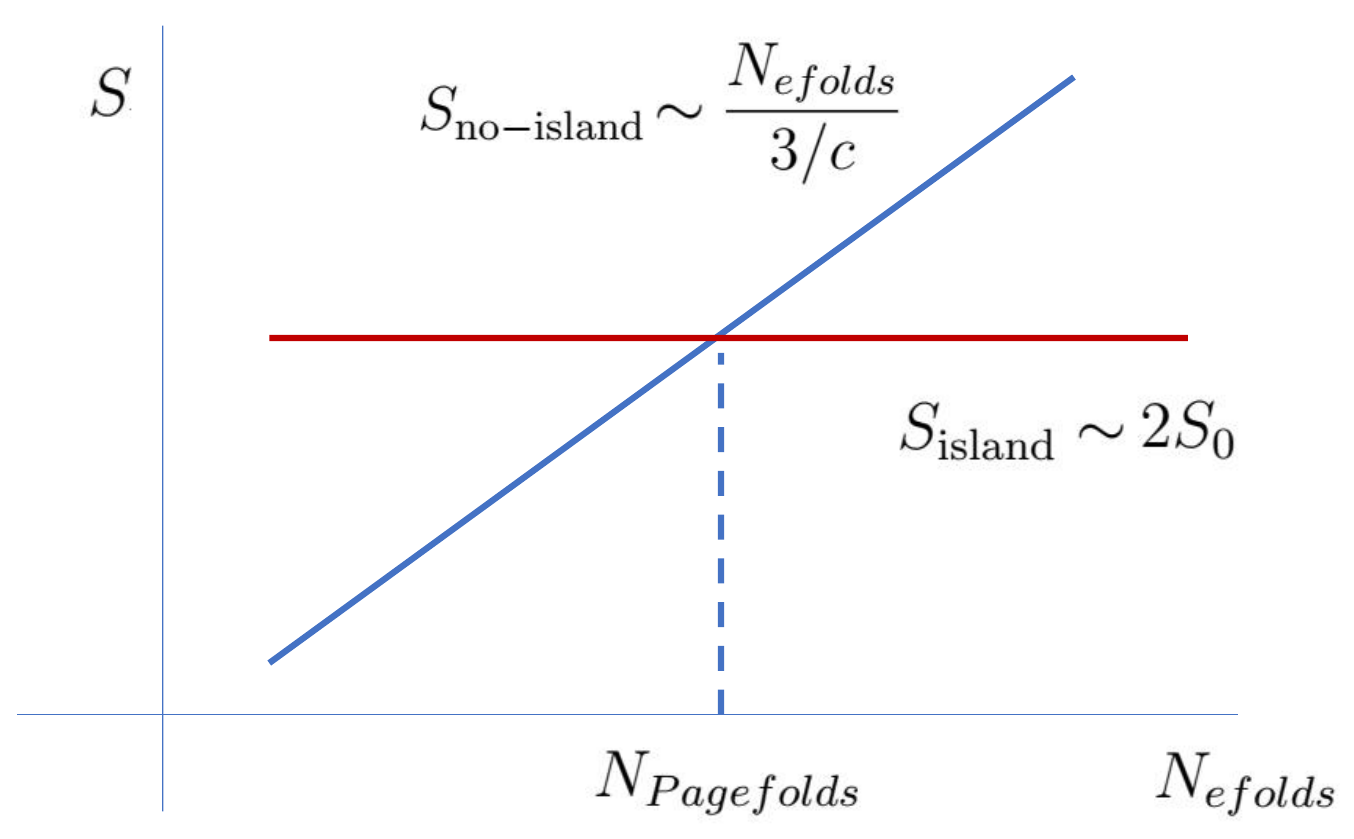}
\caption{$S_{\textsf{no-island}}$ (blue line) will exceed the dS
entropy at $N_{efolds}>N_{Pagefolds}$, which suggests a black
hole-like information paradox. However, the paradox can be
resolved by the red line, which is that with an island. In light
of the island rule, at large $N_{efolds}$, the states of
perturbation modes accessible to an observer inside a Minkowski
hat will be $2S_0$. } \label{Fig.2}
\end{center}
\end{figure}

Thus calculating $\partial S_{gen}/\partial X_I=\partial
S_{gen}/\partial T_I=0$, we have $X_I = -X_R$ and \be \coth
T_I/H^{-1} - tanh{ \lf(\sum_{p=R,I}T_p/H^{-1}\over 2 \rt)} +
\frac{6\phi_r}{c \left(sinh^2 (T_I/H^{-1})\right)} = 0\,,
\label{TIe}\ee so \be\label{islandtime} T_I \approx -{1\over
2H}\sinh^{-1} (-\frac{12\phi_r}{c}), \ee while $T_R\simeq 0$. Thus
with (\ref{islandtime}), we have \ba S_{\textsf{island}}(R) &=&
2S_{dS_4}+2\phi_r coth(T_I/H^{-1})+\frac c 3
\log\frac{2+2\cosh(T_I/H^{-1})}{\epsilon_{uv}
\epsilon_{rg}\sinh(T_I/H^{-1})(-T_R/H^{-1}) } \nonumber\\&\simeq &
2S_{dS_4}+{\cal O}(\phi_r). \label{Sisland}\ea
Thus in light of the island rule, the observer inside the
Minkowski hat collecting a number of primordial perturbation modes
will observe not (\ref{Sno}) but $S_{\textsf{island}}(R) =
2S_{dS_4}$, i.e. a Page curve, see Fig.\ref{Fig.2}. The result is
similar to that for the eternal black hole
\cite{Almheiri:2019yqk}.


The Pagefolds $N_{efolds}$ (analogous to that for black hole) also
can be calculated out. Requiring $S_{\textsf{island}}(R) =
S_{\textsf{no-island}}(R)$, we have \be N_{Pagefolds}\approx
\left({3\over c}\right)2S_{dS_4}. \label{Pagefolds}\ee
Thus 
the information paradox argued in Ref.\cite{Arkani-Hamed:2007ryv}
is resolved \footnote{see also Ref.\cite{Teresi:2021qff} for the
possible resolution in pure dS without singular patch. }. The
slow-roll inflation is able to last longer
than $N_{efolds}={S_{dS_4}}$ (though the state of independent
modes he detected is bound by $e^{S_{dS_4}}$). Here, $1\ll c\ll
\phi_r\ll S_{dS_4}$, so \be N_{Pagefolds}\ll S_{dS_4}.\ee


\subsection{``Page-like" curve for the spectrum of primordial perturbations}

However, it seems to still have a question to be solved. It has
been argued also in Ref.\cite{Arkani-Hamed:2007ryv} that in such a
patch with a period of slow-roll inflation, the observer will not
be able to detect the perturbation modes with
$N_{efolds}>S_{dS_4}$, since such modes have the amplitude
$P_{\zeta}\gtrsim 1$ so that he will be swallowed rapidly by the
black hole. How to explain it?


In Fig.\ref{Fig.3}, inside our patch, the inflaton slowly roll
along its potential III. It is well-known that the evolution of
perturbation mode $\delta\phi$ \footnote{It is not the dilaton
$\phi$ in (\ref{S}) but the scalar perturbation mode in 4D
spacetime.} is \be a(-T)\delta \phi_{k}=\lf({-\pi T\over
4}\rt)^{1/2} H_{\nu}^{(1)}(-kT)\,, \ee where $\nu\approx 3/2$ for
slow-roll inflation and $a(-T)={1\over (-HT)}$ \footnote{The
Hubble rate inside bubble is usually slightly lower than that of
parent dS region. Here, it is assumed that both are approximately
equal.}.
In the limit $-kT\ll1$ (equivalently the region $R$), we have \be
P^{\delta\phi}_{\textsf{no-island}}(R)={k^3\over
2\pi^2}|\delta\phi_{k,R}|^2=\left({H\over 2\pi}\right)^2.
\label{Pdeltaphi}\ee Thus the metric perturbation is \be P_{{\cal
\zeta},R}\sim {H^4\over {\dot\phi}^2}.\label{PR1}\ee During the
slow-roll period, \be S_{dS_4}\simeq \int {{\dot H}\over
H^4}{dN_{efolds}}\sim { N_{efolds}\over
P_{\zeta,R}}.\label{dSN}\ee Thus if $N_{efolds}=S_{dS_4}$, we have
\be P_{\zeta, R}\sim 1.\ee It seems that the perturbation modes
with $N_{efolds}> S_{dS_4}$ will inevitably have the amplitude
$P_{\zeta,R}\gtrsim 1$,
which so will swallow the observer into a black hole rapidly so
that he is able to detect only the perturbations modes with
$N_{efolds}\lesssim S_{dS_4}$.

However, actually at large $N_{efolds}$ the state of perturbation
modes at $R$ has encoded that of island in neighborly collapsed
patch, so (\ref{Pdeltaphi}) might be invalid.
In such a model, 
we will show how ${ P}_{\zeta, R}$ is possibly modified at
$N_{efolds}>N_{Pagefolds}$ scale.


\begin{figure}[t]
\begin{center}
\includegraphics[width=14cm]{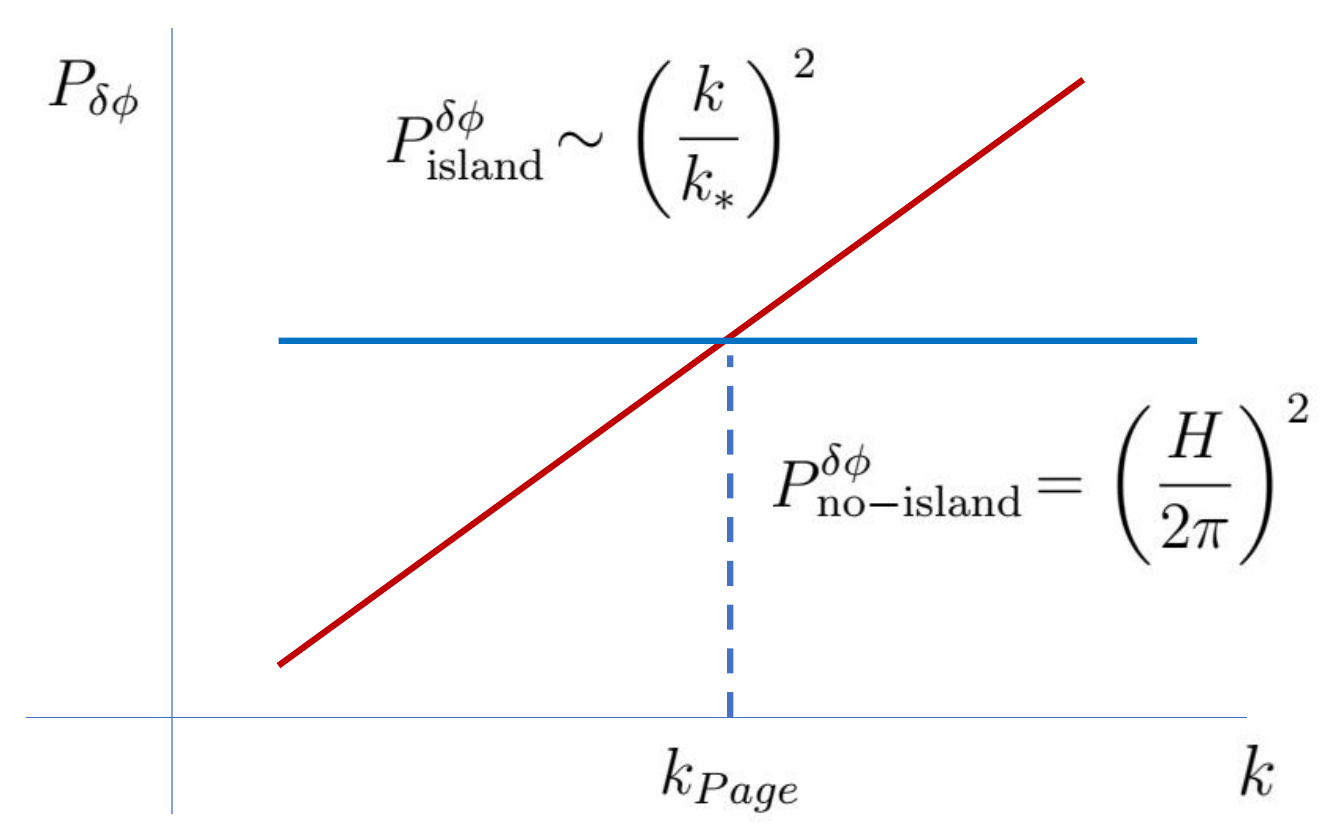}
\caption{The spectrum of primordial perturbation (blue line)
without the island is flat. However, at large scale $k<k_{Page}$,
the island appears, which modifies the spectrum and makes it
present a ``Page-like" large-scale suppression (red line). This
Fig. is similar to Fig.\ref{Fig.2} ($N_{efolds}=-\ln k$), and so
might be an observable manifestation of such an ``island". }
\label{Fig.5}
\end{center}
\end{figure}

Here, consider a pair of entangled partner, one falls into the
collapsing patch, and the other escapes into our inflating patch
and ``develop" into the perturbation mode with the wavenumber $k$,
see Fig.\ref{Fig.4}.
The state of such a system is $|\Psi\rangle\sim \sum
|\delta\phi\rangle_I |\delta\phi\rangle_R$
\cite{Penington:2019kki}.
Thus when the island is present, we might have \be {
P}^{\delta\phi}_{\textsf{island}}(R)={k^3\over
2\pi^2}\lf|\Psi_k\rt|^2\sim P^{\delta\phi}_{\textsf{
no-island}}(R)\lf(\int \lf|\delta\phi_{k,I}\rt|^2 d^3\vec{k}\rt),
\label{Pent}\ee where $|\Psi_k|^2\sim \lf(\int
\lf|\delta\phi_{k,I}\rt|^2 d^3\vec{k}\rt)|\delta\phi_{k,R}|^2$ for
a fixed wavenumber $k$, and $|\delta\phi_{k,I}|$ is the
(perturbation mode) state at $I$ with the wavenumber $k$.

As an estimate, it might be imagined that inside the collapsing
patch the initial state of $\delta\phi_k$ is ${1\over
a\sqrt{2k}}e^{-ikT}$. Regarding the collapsing bubble as a
contracting AdS universe \footnote{Inside such collapsing patches,
the evolution of spacetime might be more complicated, and consist
of multiple phases with different $w$, see Appendix B.} with the
state equation $w\gg 1$, we have \cite{Piao:2004jg,Piao:2004uq}
\be a(-T)\delta\phi_k=\lf({-\pi T\over 4}\rt)^{1/2}
H_{1/2}^{(1)}(-kT)\,.\label{aTcol}\ee Here, it is speculated that
before hitting the singularity, all modes falling into the
collapsing patch, which are entangled with the partner modes
escaped into our inflating patch, must evolve with (\ref{aTcol}),
or see Appendix B. In the limit $-kT_I\ll1$ (noting the island is
near singularity), we have \be P_{\delta\phi}(I)={k^3\over
2\pi^2}|\delta\phi_{k,I}|^2=\left({\Lambda_I\over 2\pi}\right)^2
\left({k\over k_*}\right)^2, \ee where $\Lambda_I={1\over
T_{singular}-T_I}={1\over -T_I}$. and $k_*$ is the critical
wavenumber that the effect of ``island" on the primordial
perturbations can not be neglected.

According to (\ref{Pent}), since the state of island must be
encoded in $R$, the observer able to access enough perturbation
modes would ``see" \be { P}^{\delta\phi}_{\textsf{island}}(R)\sim
\lf({k\over k_*}\rt)^2,\label{Pisland}\ee
suppressed at $k\ll k_*$ scale. Accordingly, with such an
``island" near the neighbourly singularity, the observer inside
the Minkowski hat will be able to safely access to the
perturbation modes with $N_{efolds}>S_{dS_4}$.

Though with the island, we have a large-scale suppressed
perturbation spectrum, at small scale $P_{\zeta, R}$ is still flat
\footnote{It should be mentioned that the perturbation spectrum
without island can be nearly scale-invariant $n_s-1=-{\cal
O}(0.04)$. However, $n_s=1$ is also observationally favored
\cite{Ye:2021nej,Jiang:2022uyg,Smith:2022hwi,Jiang:2022qlj,Giare:2022rvg}
in light of recent Hubble tension. In $V_{inf}\sim \phi^p$
inflation models, the slow-roll inflation might happen to end
prematurely at $\epsilon\ll 1$ so that we will have
$|n_s-1|\lesssim {\cal O}(0.001)$, see
\cite{Kallosh:2022ggf,Ye:2022efx}.}, i.e. \be P_{\zeta, R}\simeq {
H^2\over \dot\phi^2} P^{\delta\phi}_{ \textsf{no-island}}(R)\sim
{H^4\over {\dot\phi}^2}. \ee Thus the spectrum of primordial
perturbations that the observer inside the Minkowski hat can see
also follow a ``Page-like" curve, see Fig.\ref{Fig.5}, which might
be an ``observable" manifestation of island recovering the Page
curve for the state of perturbation modes.



\subsection{Discussion}

Here, it must be required that our patch is entangled with a
collapsed patch (or a black hole) \footnote{In the spirit of
ER=EPR \cite{Maldacena:2013xja}, both entangled but disconnected
patches might actually be connected with wormholes.} so that the
state of island inside the collapsed patch is encoded in our
primordial perturbations.
Thus it seems that in such an inflating multiverse, the primordial
perturbations in our observable Universe might be nothing but the
Hawking radiation of the collapsed singularity or black hole, see
Fig.\ref{Fig.4}.

\begin{figure}[t]
\begin{center}
\includegraphics[width=15cm]{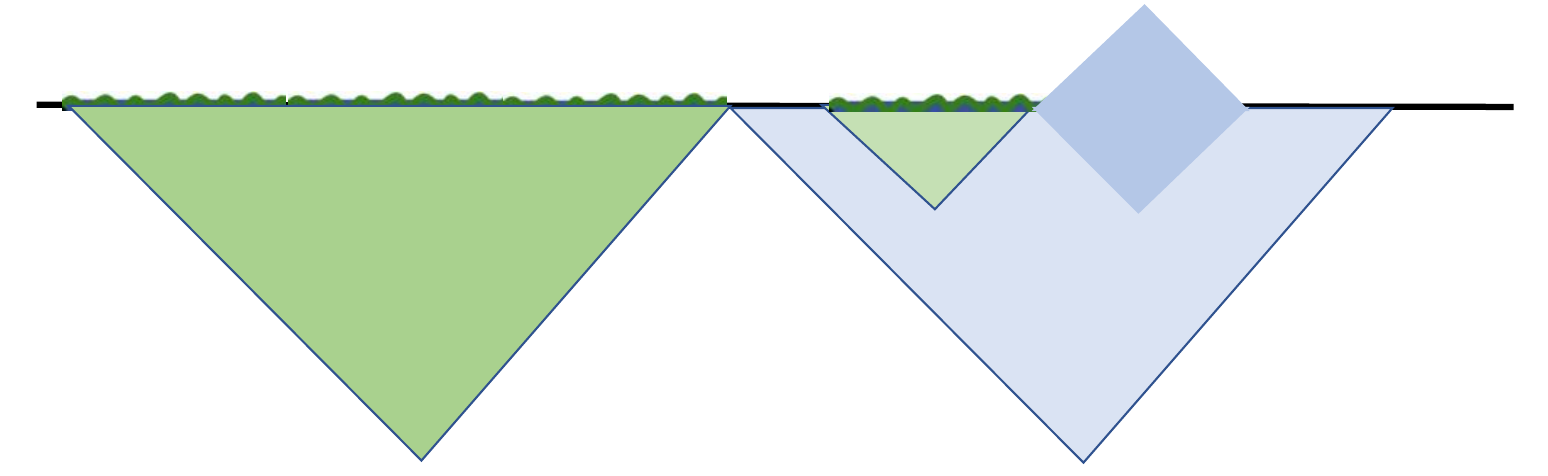}
\caption{It might be that a local dS patch (blue) and a collapsed
patch (green) constitute an entangled pair, their parent dS region
and another collapsed region also constitute an entangled pair,
and so on, so that we have an eternally inflating multiverse. }
\label{Fig.7}
\end{center}
\end{figure}

It is usually conjectured that a breakdown of the dS effective
field theory will occurs at $t\sim {S_{dS_4}\over H}$, which so
suggests the bound (\ref{dSbound}), i.e. $N_{efolds}\lesssim
S_{dS_4}$. Here, we showed that such a bound might be nonexistent,
while the state of primordial perturbation modes the observer is
able to see follows a Page curve, however, such an inflating patch
must be accompanied with a neighbourly collapsed patch
\footnote{However, if the slow-roll inflation last only a short
period, i.e. $N_{efolds}\ll S_{dS_4}$, we might not need such a
pair of entangled patches.} so that both patches constitute an
entangled pair. It is also possible that accompanied with a
partner-like collapsed region, parent dS region (both our patch
and the collapsed patches nucleated) might also have unbounded
$N_{efolds}$, and so on, see Fig.\ref{Fig.7}. It will be
interesting to investigate whether such a result helps to
understand the holographic descriptions for eternal inflation
\cite{Susskind:2007pv,Sekino:2009kv,Vilenkin:2013loa,Hartle:2016tpo,Nomura:2016ikr},
see e.g.recent \cite{Langhoff:2021uct}


It is also interesting to note that if $N_{Pagefolds}\approx 60$,
the large-scale suppression of primordial perturbation spectrum
\footnote{In bounce inflation models
\cite{Piao:2003zm,Piao:2005ag,Liu:2013kea,Qiu:2015nha,Cai:2016thi,Cai:2017pga},
similar large-scale suppression also appeared. However, here we
have not a ``bounce", it is the entanglement that makes the state
inside the collapsed patch is encoded in the primordial
perturbations in our patch.} will happen at low-$l$ scale of CMB
in our observable Universe. Thus it might be expected that such
large-scale anomalies (so the effect of ``island"
\cite{Almheiri:2019hni} or replica wormhole
\cite{Penington:2019kki,Almheiri:2019qdq}) will have significant
observable imprints in CMB, so that we might have the opportunity
to see the ``island" and the Page curve in the sky. Here, our
result for the spectrum of primordial perturbations at
beyond-Pagefolds scale is only speculative, and the right spectrum
recording the ``island" effect might be more complicated, which,
however, upcoming high-precision CMB observation might tell us.
According to (\ref{Pagefolds}), for the inflation occurring at
$10^{16}$Gev, we have \be c\simeq{S_{dS_4} \over N_{Pagefolds}}
\sim 10^{10},\ee consistent with $c\gg 1$.


\section*{Acknowledgments}

This work is supported by the NSFC, No.12075246 and the
Fundamental Research Funds for the Central Universities.

\appendix

\section{de Sitter JT spacetime }

\begin{figure}[t]
\begin{center}
\includegraphics[width=14cm]{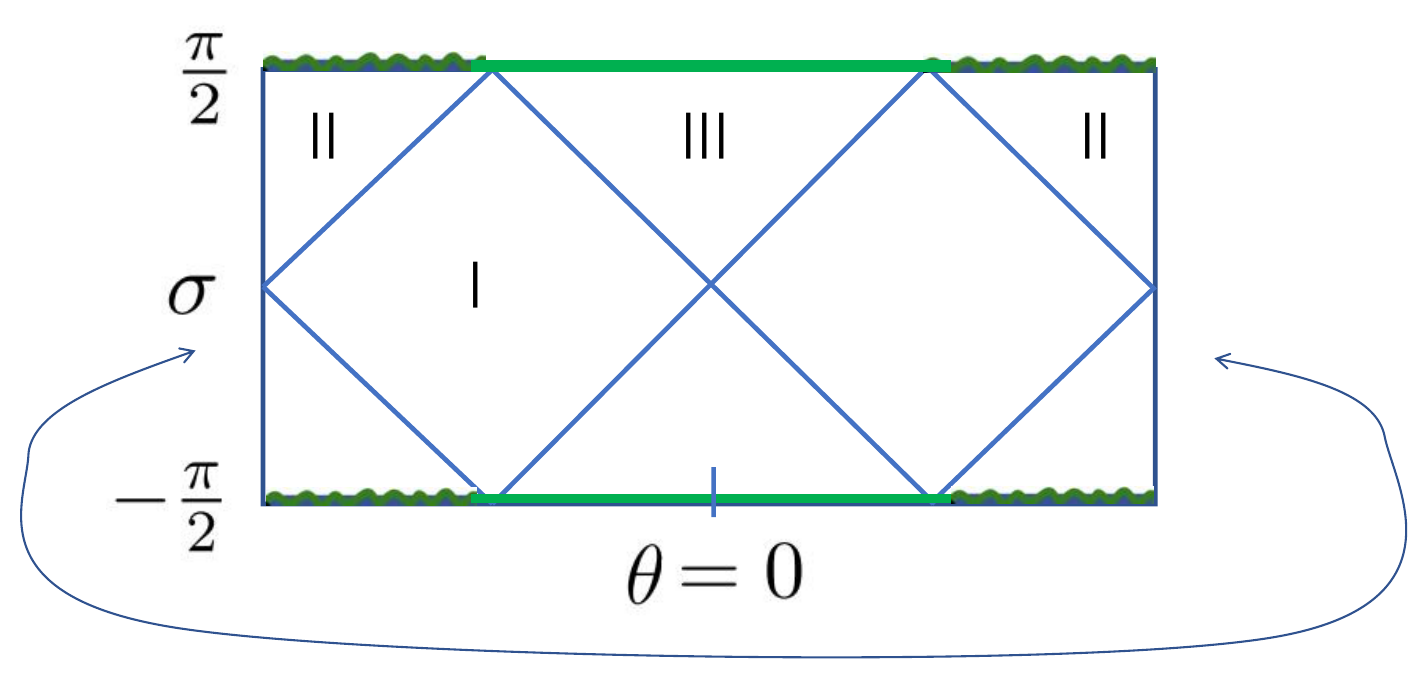}
\caption{de Sitter JT spacetime. The patch II and III correspond
to the collapsing patch and the expanding dS
patch, respectively. 
  }
\label{Fig.6}
\end{center}
\end{figure}

In the global coordinate, the metric of JT dS is (\ref{SchdS})
with $-\pi/2\leqslant \sigma \leqslant\pi/2$ and $-\pi\leqslant
\theta \leqslant \pi$. The Penrose diagram is plotted in
Fig.\ref{Fig.6}, see also \cite{Maldacena:2019cbz,Cotler:2019nbi}.

In patch III, the dilaton approaches $+\infty$ as $\sigma=\pi/2$
(the ${\cal I}^+$ boundary). In patch II, the dilaton approaches
$-\infty$ as $\sigma=\pi/2$ (a black hole-like singularity). Thus
JT dS spacetime can be thought as a nearly-dS expanding patch with
the collapsing patches (or black holes) on the left and right
side, respectively. 

In the conformal complex coordinate $x^{\pm} = e^{-i(\sigma \pm
\theta)}$, we have a Weyl-equivalently flat metric $ds^2 ={dx^+
dx^-\over \Omega^{2}}$. Thus the 2D CFT entanglement entropy at
$R$ in Fig.\ref{Fig.1} is \be S_{mat}(R) = \frac {c} {6} \log
\frac{\Delta x^- \Delta
x^+}{\epsilon_{uv}^2\Omega(x_1)\Omega(x_2)}=\frac {c}
{6}\log{2\cos(\sigma_1-\sigma_2)-2\cos(\theta_1-\theta_2)\over
\epsilon_{uv}^2\cos\sigma_1\cos\sigma_2}, \label{CFT}\ee where the
interval has endpoints at $x_1$ and $x_2$. (\ref{CFT}) can be
rewritten as (\ref{SmatR}) with \be ctan\sigma= -{\sinh
(T/H^{-1})\over \cosh X/H^{-1}},\quad\quad ctan\theta={\cosh
(T/H^{-1})\over \sinh X/H^{-1}}. \label{ctan}\ee




\section{On perturbation modes inside AdS collapsed patch}

Inside the AdS bubble the collapsing evolution can hardly be
described by a single state equation $w=const.$, since different
stage of collapse might correspond to different $w$.

As an estimate, we consider that the collapsing evolution consists
of different phases $w_j=const.$. The perturbation equation of
$\phi$ inside the collapsing patch is \be u_k^{\prime
\prime}+\left(k^{2}-\frac{a^{\prime \prime}}{a}\right)
u_k=0\,,\label{MS-eq02} \ee where $u_k=a\delta\phi_k$.
The different phases of collapse are signed with
$\epsilon_j=-{\dot H}_j/H^2_j={3\over 2}(1+w_j)$. It has been
showed in Ref.\cite{Cai:2015nya} that for the $j$-th phase, we
have \be a_{j}\sim
\left({T}_{*,j}-T\right)^{\frac{1}{\epsilon_{j}-1}}, \ee where $
{T}_{*,j} =T_{j}- {1\over (\epsilon_{j}-1) a(T_j)H(T_j)}$ is set
by requiring the continuity of $a$ at the end of phase $j$ (i.e.,
$T=T_j$). Thus we have \be \frac{a_j^{\prime
        \prime}}{a_j}={\nu_j^2-{1/4} \over
    {(T-T_{*,j})^2}}\,,\label{zppbz02} \ee where
$\nu_j={3\over2}\lf|{1-w_j\over 1+3 w_j}\rt|$.
Regarding the phases $j$ and $j+1$ as adjacent phases, we have the
solutions to Eq.(\ref{MS-eq02}) as \ba u_{k,{j+1}}&=&{\sqrt{\pi
(T_{*,{j+1}}-T)}\over
    2}\Big\{\alpha_{j+1}H_{\nu_{j+1}}^{(1)}[k(T_{*,j+1}-T)]
\nn\\
&\,&\qquad\qquad\qquad\quad
+\beta_{j+1}H_{\nu_{j+1}}^{(2)}[k(T_{*,j+1}-T)] \Big\} ,\qquad
T_{j}<T<T_{j+1}\,, \label{solution}\ea where $\alpha_{j+1}$ and
$\beta_{j+1}$ are $k$-dependent coefficients. Thus with the
matching conditions $u_{k,j}(T_{j+1})=u_{k,j+1}(T_{j+1})$ and
$u_{k,j}'(T_{j+1})=u_{k,j+1}'(T_{j+1})$, we have \ba \left(
\begin{array}{ccc} \alpha_{j+1}\\ \beta_{j+1}
\end{array}\right)
&=& {\cal M}^{(j)}
\left(\begin{array}{ccc} \alpha_{j}\\
    \beta_{j}\end{array}\right)\,,\quad {\rm where} \quad
{\cal M}^{(j)} = \left(\begin{array}{ccc}
    {\cal M}^{(j)}_{11}&{\cal
        M}^{(j)}_{12}\\
    {\cal M}^{(j)}_{21}&{\cal M}^{(j)}_{22}\end{array}\right)\,,
 \label{Mmetric}
\ea see Refs.\cite{Cai:2015nya,Cai:2019hge} for ${\cal M}^{(j)}$.
The information of $j=1,2\cdots j$ phases has been encoded fully
in the Bogoliubov coefficients $\alpha_{j+1}$ and $\beta_{j+1}$.

As an example, we consider a model in which the collapse consists
of the three ($j=1,2,3$) phases. Thus we have \ba
u_{k,3}&=&{\sqrt{\pi (T_{*,3}-T)}\over
2}\Big\{\alpha_{3}H_{\nu_3}^{(1)}[k(T_{*,3}-T)]\nn\\
&\,&\qquad\qquad\qquad\quad+\beta_{3}H_{\nu_3}^{(2)}[k(T_{*,3}-T)]
\Big\}.\ea 
The resulting spectrum of perturbation will be \be
P_{\delta\phi}=\frac{k^3}{2\pi^2
}\left|\delta\phi_k\right|^2=2^{2\nu_3-3}{\Gamma^2(\nu_3) }
{\lf(T_*-T\rt)^{1-2v_3}\over \pi^3a^2k^{2\nu_3-3}}\lf|\alpha_3 -
\beta_3 \rt|^2\,,\label{eq:PT} \ee \ba \left(
\begin{array}{ccc} \alpha_{3}\\ \beta_{3}
\end{array}\right)
&=& {\cal M}^{(2)} {\cal M}^{(1)}
\left(\begin{array}{ccc} \alpha_{1}\\
    \beta_{1}\end{array}\right)\,.
\label{Mmetric-2} \ea  
In particular, considering $w_3\gg 1$ for the $j=3$th phase
($\nu_3=1/2$) and the Bunch-Davis initial state $u_k=
\frac{1}{\sqrt{2 k} }e^{-i kT}$ (so $|\alpha_1|=1$ and
$|\beta_1|=0$), we have \be P_{\delta\phi}\sim
k^2\lf|\alpha_3-\beta_3\rt|. \ee It is noted that if the collapse
consists of only one single phase with $w\gg 1$ (the field is
rolling in AdS well and ${\dot\phi}^2\simeq |V(\phi)|$), we have
\be P_{\delta\phi}\sim \lf({k\over k_*}\rt)^2, \ee and with $w=1$
(the field climbed out of the AdS well and $\dot\phi^2\gg V$
e.g.\cite{Piao:2004hr,Piao:2004me}), we have \be
P_{\delta\phi}\sim \lf({k\over k_*}\rt)^3.\ee However, it is also
interesting to note that if $w=1/3$ (the collapsing spacetime
contains only radiation or massless particles), we also have
$P_{\delta\phi}\sim {k^2}$, see also
\cite{Piao:2004jg,Piao:2004uq}.

\end{document}